\documentclass[referee]{aa} 

\usepackage{graphicx}
\usepackage{txfonts}

\begin{document}

   \title{Modeling of CoRoT and Spitzer lightcurves in NGC 2264
              caused by an optically thick warp}


   \author{E. Nagel
          \inst{1,2}  
          \and
          J. Bouvier\inst{2}
          }

   \institute{Departamento de Astronomia, Universidad de Guanajuato, Mexico\\
	      \email{erick@astro.ugto.mx}
	 \and
	     Univ.Grenoble Alpes, IPAG, 38000 Grenoble, France\\
	     \email{jerome.bouvier@univ-grenoble-alpes.fr}
	     }		

   \date{Received 28 July, 2018; accepted 19 March, 2019}

 
  \abstract
   {}
   {We present an analysis of simultaneously observed CoRoT and Spitzer lightcurves for $4$ systems in the stellar forming region NGC 2264: Mon-660, Mon-811, Mon-1140 and Mon-1308. These objects share in common a high resemblance between the optical and infrared lightcurves, such that the mechanism responsible to produce them is the same. The aim of this paper is to explain both lightcurves simultaneously with only one mechanism.  }
   {We have modeled the infrared emission as coming from a warp composed of an optically thick wall and an optically thick asymmetric disk beyond this location. We have modeled the optical emission mainly by partial stellar occultation by the warp. }
   {The magnitude amplitude of the CoRoT and Spitzer observations for all the objects can be described with the emission coming from the system components. The difference between them is the value of the disk flux compared with the wall flux and the azimuthal variations of the former. This result points out the importance of the hydrodynamical interaction between the stellar magnetic field and the disk.} 
   {CoRoT and Spitzer lightcurves for the stellar systems Mon-660, Mon-811, Mon-1140 and Mon-1308 can be simultaneously explained using the emission coming from an asymmetric disk and emission with stellar occultation by an optically thick wall.}


   \titlerunning{Modeling of CoRoT and Spitzer lightcurves}

   \maketitle
%

\section{Introduction}

Young stellar objects (YSOs) flux variability is commonly detected in the optical and in the infrared. For the young stellar cluster NGC 2264 there are several observational studies that clearly show variability for a large fraction of the observed objects; in the optical with the CoRoT Space Telescope (Alencar et al. 2010, Cody et al. 2014, Stauffer et al. 2015,2016) 
and in the infrared with the Spitzer Space Telescope (Morales-Calder\'on et al. 2011,Cody et al. 2014, Stauffer et al. 2015,2016). Stauffer et al. (2016) focus on a sample of stellar stochastic lightcurves that can be explained with changes on $\dot{M}$ that leads to a variable dust heating of the material rotating around the object. Some of the objects show the IR and optical lightcurves resembling each other, suggesting that the physical mechanism responsible for their shape is the same. 

In NGC 2264, Alencar et al. (2010) use CoRoT lightcurves to search for a subsample with a periodic signal that can be explained with  occultations by an inner warp as AA Tau (Bouvier et al. 1999). They conclude that at least $\sim 30$ to $40\%$ of YSOs with inner dusty disks present this kind of behavior. In the Orion Nebula Cluster, Morales-Calder\'on et al. (2011) present Spitzer lightcurves at $3.6$ and $4.5\mu$m of 41 objects that show flux drops with duration of one to a few days that can be interpreted with material crossing the line of sight. From this set, they extract one third with a detected periodic dip, pointing out that structures moving at a keplerian angular velocity are obvious features shaping the lightcurves. For this same region, Rice et al. (2015) confirm that 73 out of a sample of 1203 objects present AA Tau-type periodic lightcurves. 

Bouvier et al. (1999) suggest that such structures can be associated with the accretion of material along the stellar magnetic field lines, where the magnetic dipolar axis is inclined with respect to the disk rotational axis. The magneto-hydrodynamical (MHD) simulations of this system configuration by Romanova et al. (2013) show besides the magnetospheric streams an uniformly rotating bending wave located between the location of the streams and the outer vertical resonance. Both components can shadow the star when the system is seen at high inclination, thus becoming a physical mechanism to explain the lightcurves shape. Theoretical analysis of the MHD equations by Terquem $\&$ Papaloizou (2000)also arrives to the conclusion that a shadowing warp is formed. 

Inner disk structure can explain azimuthal surface brightness asymetry in the object TW Hya that moves at a constant angular velocity consistent with shadowing material rotating at a keplerian velocity associated to a radius around $1$AU (Debes et al. 2017). Also inhomogeneities very close to the star are responsible for the shape of the optical and IR lightcurves in the young low-mass star ISO-Oph-50 (Scholz et al. 2015). From the observations of the IC 348 cluster Flaherty et al. (2012) point out 3 low luminosity objects (LRLL 58,67,1679) with non-periodic lightcurves in the near-IR that can be interpreted with dust that moves along the stellar magnetic field lines.

From this, we can conclude that the material distributed asymmetrically in the innermost region of the disk is relevant to explain the variability of the lightcurves either in the optical or in the IR, as can be seen from the aforementioned observational studies and on the theoretical works that sustain this information. In this work on the interpretation of the variability of the lightcurves we point out for each object the importance of the disk flux compared with the wall flux and the amplitude for the azimuthal contribution of the former. This allows to qualitatively characterize the innermost disk structure that we cannot resolve with current instrumentation. This characterization is important because there are many observed planets located very close to the star and their evolution towards this location strongly depends on the physical conditions of the disk inner part in the initial stage of the system life. 

	McGinnis et al. (2015) studied the photometric variability of young stellar
objects in the star forming region NGC 2264. They present CoRoT lightcurves for 33 objects showing AA Tau-type behavior: the optical lightcurve can be described with periodical stellar occultations due to an optically thick warp.
From this set, McGinnis et al. (2015) present simultaneous   observations of 29 stars in the optical and the infrared using the CoRoT and the Spitzer Space Telescope. 
A comparison between the lightcurves at both wavelength ranges show different behaviors. In this study, we will focus on the 4 objects that present AA Tau-like modulation in the IR and in the optical which closely follows each other, meaning that the mechanisms that are generating the variability in the optical and in the IR are related. The objects are: Mon-660, Mon-811, Mon-1140 and Mon-1308. 

The aim of this work is to consistently reproduce the observed amplitude of the CoRoT magnitude $\Delta [CoRoT]$, along with the observed amplitudes in the IR: $\Delta [3.6]$ and $\Delta [4.5]$. The material that it is occulting the star and which is responsible for the changes in $[CoRoT]$ is the one that is producing the excess in the IR range, thus responsible for changes at $[3.6]$ and $[4.5]$microns.

A parametric study about the effect of a dust distribution around a star in optical and IR lightcurves was previously done by Kesseli et al. (2016). They used a Monte Carlo radiation transfer code described in Whitney et al. (2013) that includes heating by stellar radiation and by accretion. Besides the dust emission, Kesseli et al. include hotspots at different latitudes to produce different amounts of magnitude variability due to stellar rotation. Their models qualitatively reproduce the periodic dippers analysed in McGinnis et al. (2015) using a disk warp that changes in radius and in azimuthal angle (see Eq. 8 in Whitney et al. 2013) in a way similar to the 2D warp shape used in Bouvier et al. (1999) for the interpretation of the optical lightcurve of AA Tau. 
Our idea is to follow the same analysis but applied to specific objects. 

In Section~2 we present the description of the modeling, followed in Section~3 by the results of the lightcurves modeling for the set of 4 YSOs, Section~4 contains the discussion and finally Section~5 gives the conclusions.

\section{Modeling}
\label{Modeling}

\subsection{Main aspects of the modeling}
\label{Aspects}

The disk surrounding the star has two main components: an optically thick vertical wall and an asymmetrical emitting structure beyond this location. These two components merge to form the warp (see Figure~1 for a sketch of the system). 

   \begin{figure}
   \centering
   \includegraphics[width=\hsize]{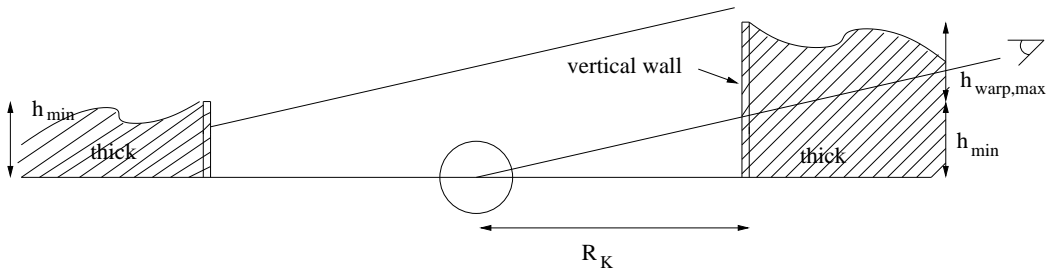}
      \caption{Sketch of the system modeled. 
The inclined straight lines represent two lines of sight with the same inclination crossing different sections of the system. The left line contributes to emission of the wall and the right line contributes to emission of the disk. In this case, all the stellar surface is occulted.  
              }
         \label{sketch}
   \end{figure}

The optically thick vertical wall is located at the Keplerian radius, 
$R_{K}$, which is consistent with the observed period of the lightcurve. We assume that the main mechanism shaping the wall is magnetospheric accretion at the magnetospheric radius ($R_{mag}$) such that $R_{K}\sim R_{mag}$. 
Thus, it is reasonable to assume the presence of a wall, and, as a zero order approach, we consider a stationary disk configuration which it is perturbed to form the vertical wall. At the wall location, we expect a highly dynamical environment due to the MHD interaction between the disk and the stellar magnetosphere such that the stationary disk configuration is not reasonable. In the context of this work, this configuration is used to give characteristic values to the density and the vertical distribution of grain sizes. These values are required to get the wall temperature ($T_{wall}$) but it mainly depends on the distance to the star, thus, the values taken do not significantly change the modeled wall emission. In Figure~2 we present $T_{wall}$ in terms of height for the stellar objects Mon-660, Mon-811 and Mon-1308. We do not include the plot for Mon-1140 because this object has not a dusty wall (see Section~3.3 for details).
We expect in reality that the material distribution in this region is much more complex than the toy model used here as it is shown in the 
analysis of the young low-mass star ISO-Oph-50, where the modeling by Scholz et al. (2015) requires inhomogeneties in the inner disk that can be the result of a turbulent environment. In any case, we also think that an optically thick structure is responsible for the main features of the lightcurves analyzed in this paper. 
  
   \begin{figure}
   \centering
   \includegraphics[angle=-90,width=\hsize]{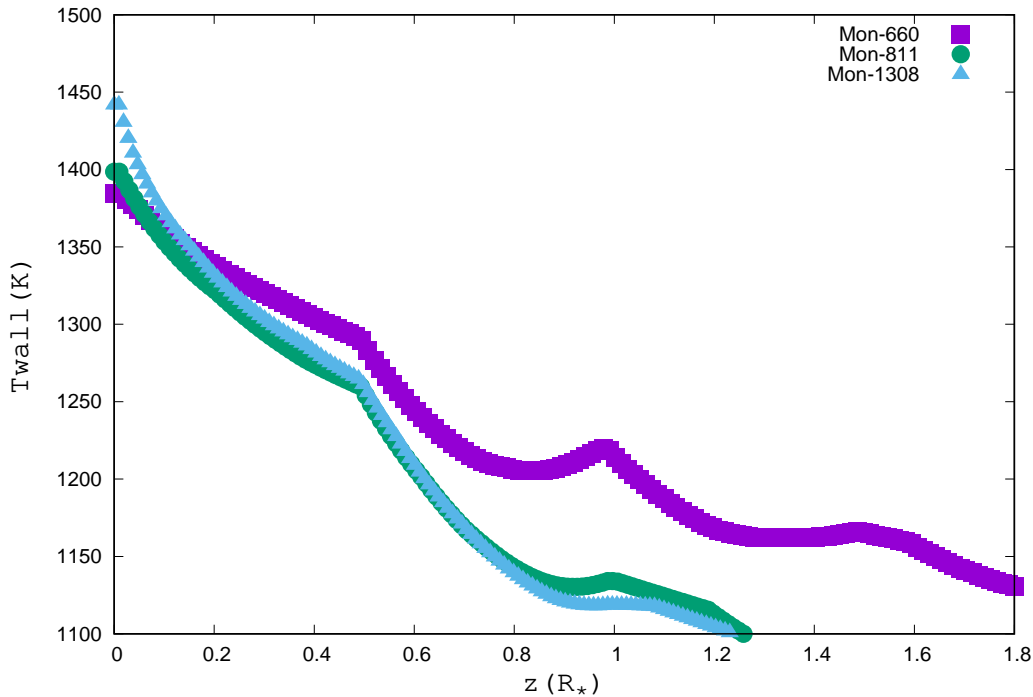}
      \caption{The wall temperature ($T_{wall}$ in Kelvin) at the vertical wall as a function of height ($z$ in stellar radius) for Mon-660, Mon-811 and Mon-1308}
         \label{Twall-Mon}
   \end{figure}

There is dust inside the magnetosphere when the sublimation radius $R_{sub}<R_{mag}$. Due to the stellar magnetic field lines, this dust is not located in a stationary disk, thus it is not easily characterized. In terms of the modeling, the emission coming from the material (gas plus dust) inside 
the magnetosphere and from the outer disk is set by the flux required to explain the observed IR flux and the amount of variability in the Spitzer lightcurves. Emission from gas inside the sublimation radius is suggested in the interferometric observations of Herbig Ae stars MWC 275 and AB Aur (Tannirkulam et al. 2008). In order to explain the observations, the flux should account for 40-60$\%$ of the total K-band emission. Akeson et al. (2005) found the same result for the T Tauri star RY Tau. However, McClure et al. (2013) conclude from the fitting of NIR emission of a sample of T Tauri stars that there is not evidence of emission from optically thick gas inside the sublimation radius.

In order to confirm the existence or not of dust inside the magnetosphere we calculate $R_{sub}$ as in Nagel et al. (2013) assuming that the large grains are located in the disk midplane. In the boundary between the dusty and dust-free disk, there is a large opacity change such that it is formed an optically thick structure. The temperature of this structure ($T_{wall}$) is calculated starting from the stellar radius and increasing the value up to the radius where $T_{wall}=T_{sub}$. The radius where this condition is fulfilled is $R_{sub}$. $T_{wall}$ is calculated with the analytical expression given in Nagel et al. (2013). $T_{sub}$ is a function of gas density and it is taken for different grain species from Table 3 in Pollack et al. (1994). The gas density is taken assuming as typical the density of a stationary configuration for a settled disk. The density structure is taken using the codes in D'Alessio et al. (1998). The dust is distributed in the disk using two grain size distributions: one for large grains close to the midplane and another for small grains in the upper layers. The transition between these two distributions is modeled as in D'Alessio et al. (2006). 

As we fully explain in Section~2.3, the height of the vertical wall is fixed by estimating the amount of occultation required to explain the CoRoT lightcurves. Physically this is possible because the highly dynamical interaction in this region allows to move the large grains located in the midplane layer in the stable state (turning off the magnetic stellar field) towards larger heights. Also note that this is in accord with the interpretation of the observations of Mon-660 by Schneider et al. (2018), where in order to interpret some optical spectra, an optically thick wall is required that produces complete occultation of a fraction of the stellar surface $f_{C}$, an optically thin wall that extinct another fraction of this surface $f_{B}$ and finally an unocculted fraction $f_{A}$: $f_{A}+f_{B}+f_{C}=1$. Besides, Schneider et al.'s (2018) model also requires an increment on the abundance of dust grains in the optically thin layer that can be obtained with the
  dust that arrives due to the hydrodynamical perturbation.  

For Mon-660, Schneider et al. (2018) estimate the gas density in the extincted region, which is above the optically thick wall. They use an analysis of the Na and K optical doublets to derive an absorbing column density for these species, which corresponds to a hydrogen density equal to $2\times 10^{19}cm^{-2}$. The disk warp is located at $0.1$AU and they assume a radial thickness along the line of sight of the same magnitude, ending with a volumetric density $n_{H}\sim 10^{7}cm^{-3}$, in terms of mass, $\rho_{extinc}=1.67\times 10^{-17}gcm^{-3}$. This density is responsible to extinct the fraction $f_{B}$ of the stellar surface. If we run models using this value for the density in a layer above the optically thick wall then we conclude that the contribution to the IR flux can be neglected. Thus, we do not include this optically thin component in our modeling of the IR lightcurves.

The grain size distribution ($n(a)$) is taken from citet{mathis} as typical for the ISM, which is extensively used in protoplanetary disks; $n(a)\sim a^{-3.5}$. The size range for the small grains located in the upper layers of the wall is: $a_{min}=0.005\mu$m and $a_{max}=0.25\mu$m. For the larger grains located close to the midplane, $a_{min}=0.005\mu$m and $a_{max}=1$mm. The dust mass fractions compared to the gaseous mass are $\zeta_{sil}=0.004$ and $\zeta_{grap}=0.0025$ for the silicate and the graphite components, respectively. 

The MHD simulations by Kulkarni \& Romanova (2013) show that the perturbation on the disk is not only located close to $R_{mag}$ but can move further out on the disk. The structure of this region is highly complex but for the sake of the simple modeling presented here, this outer zone contributes with an asymmetrical flux parameterized as in Section~2.2.
The MHD simulations of a tilted stellar magnetic field interacting with a disk of Romanova et al. (2013) focus on the formation of waves in the disk. They conclude that a bending wave (out-of-plane modes) is formed between the corotational resonance (located at the corotational radius $R_{cr}$) and the outer vertical resonance located at $R_{ovr}=4^{1/3}R_{cr}$. In our case, $R_{cr}=R_{k}=R_{mag}$ such that the bending wave is the structure responsible for the emission of the outer component of the warp. We estimate its emission as the missing contribution required to explain the Spitzer lightcurves (see Section~2.2). This whole structure corotates with the magnetosphere such that it is relevant to explain the periodic lightcurves.

\subsection{Emitted flux}
\label{sec:flux}

The flux observed is modeled using

\begin{equation}
	F=F_{wall}+F_{disk}+f_{A}F_{\star},
\end{equation}

where $F_{wall}$ is the flux coming from the wall, $F_{disk}$ is the flux coming from the region beyond the wall and $F_{\star}$ is the flux coming from the unocculted star.

The emission from the optically thick wall comes from its atmosphere, and the radiation is extincted with the material between this location and the observer. Each layer of the atmosphere (located at an optical depth $\tau$) is emitting as a blackbody at a temperature given as in D'Alessio et al. (2005),

\begin{equation}
	T_{wall}^{4}=\alpha {F_{0}\over 4\sigma}(C_{1}'+C_{2}'e^{-q\tau}+   
                     C_{3}'e^{-\beta q\tau}),
\end{equation}

where $\alpha=1-w$, $w$ is the mean albedo to the stellar radiation, 
$\beta=(3\alpha)^{1/2}$, $F_{0}=L_{\star}/4\pi R_{wall}^2$ is the stellar flux that heats the wall at the radius $R_{wall}$, and $\sigma$ is the Steffan-Boltzmann constant.
$C_{1}'$, $C_{2}'$,$C_{3}'$,$q$ and $w$ depend on the mean Planck opacities (absorption and scattering) one set of them calculated using the typical wavelength range of the stellar radiation and the other set calculated using the typical wavelength range of the disk radiation. The scattering of stellar radiation is assumed isotropic. Thus, the total wall flux at each frequency ($\nu$) is calculated as an integral over the solid angle $\Omega$ subtended by the wall and over the optical depth $\tau$, 
 
\begin{equation}
	F_{wall}=\int_{0}^{\tau_{wall}} \nu B_{\nu}(T_{wall})e^{-\tau}d\tau d\Omega.
\end{equation}

The emission comes from the wall atmosphere which is defined with the optical depth $\tau$, where $\tau=0$ in the surface of the wall and increases towards larger radius until $\tau=\tau_{wall}=1.5$. 

The emitted flux coming from the disk in the IR ($3.6$ or $4.5$) is parameterized as 

\begin{equation}
	F_{disk}=<F_{disk}>(1-\delta|cos(2\pi (\phi-\phi_{0})/2)|)
\end{equation}

which is consistent with the warp shape described in equation~6, where $\phi_{0}$ is the phase with the lowest flux contribution. 
Because the behavior at $3.6\mu$m and $4.5\mu$m is analogous, we focus on $4.5\mu$m when referring to IR emission, such that the aim is to explain the $4.5\mu$m Spitzer lightcurve. Physically, this corresponds to a structure that moves with the same periodicity of the wall. $\phi$ corresponds to the phase of the lightcurves, $\delta$ is a free parameter fitted to explain the amplitude of the observed $[4.5]$ ($\Delta[4.5]$) and $<F_{disk}>$ is the mean disk flux required to be consistent with the mean flux extracted from the IR lightcurves ($<F_{obs}>$), this value is fixed using the following equation

\begin{equation}
	<F_{obs}>=<F_{disk}>+<F_{wall}>
\label{eq:Fdisk}
\end{equation}

where $<F_{wall}>$ is the mean value of the IR emission coming from the wall.
The $\delta$ value parametrizes the azimuthal variation of $F_{disk}$. We associate this variation to the hydrodynamical waves formed by the interaction between the stellar magnetic field and the disk. If $\delta$ increases then the effective area of the wave that it is facing the observer increases as does the flux emitted by the structure.

\subsection{Warp geometry}
\label{sec:warp}

The optically thick part of the warp is required to block some of the stellar radiation. This structure is asymmetric with a height given by $h_{warp}$. The axisymmetric part of
the structure consists of a wall with height $h_{min}$.
The values for $h_{warp}$ are taken from the warp model of Bouvier et al. (1999) and used by Fonseca et al. (2014) for Mon-660, 

\begin{equation}
	h_{warp}=h_{warp,max}|cos({{\pi(\phi-\phi_{0})}\over {2\phi_{c}}})|+h_{min},
	\,\,\,\,\,\,\,\,\,\,\,\,|\phi-\phi_{0}|<\phi_{c},
\label{eq:hwarp}
\end{equation}
\begin{equation}
	h_{warp}=h_{min},
	\,\,\,\,\,\,\,\,\,\,\,\,|\phi-\phi_{0}|>\phi_{c},
\end{equation}

where we include the axisymmetric section. We checked that the height $h_{min}$ is not eclipsing any section of the star in order to be consistent with the models in McGinnis et al. (2015). This height also satisfies $h_{min}/R\sim 0.1$ which is a typical value for thin protoplanetary disks. Note that $h_{warp,max}+h_{min}$ is the largest warp height in the models by McGinnis et al. (2015).

When the maximum height of the warp, $h_{warp,max}+h_{min}$, is along the line of sight then the optical lightcurve reaches the largest magnitude (lowest flux). For a given inclination $i$,
one can find the value for $h_{warp,max}$ which it is necessary to explain the observed ($\Delta [CoRoT]$). $h_{warp}$ is increased by 0.05 steps in order to find $h_{warp,max}$. For all the objects, the couples are given in the Appendix~A1.

\subsection{Modeling ingredients}

We use the flux calculated as described in Section~2.2 to estimate $[CoRoT]$ and $[4.5]$. We arbitrarily set the lowest magnitude as zero for both wavelengths. Thus, $\Delta [CoRoT]$ and $\Delta [4.5]$ are calculated with respect to this point. For each pair ($h_{warp,max}$,$i$), we have two remaining free parameters that can be varied in order to explain the magnitude variability in the IR, i.e. $\Delta [4.5]$. The parameter $\delta$ (see Section~2.2) is the one used to fit $\Delta [4.5]$. Note that a larger value of $\delta$ implies a larger value for $\Delta [4.5]$, thus, we increase(decrease) the value until the observed $\Delta [4.5]$ is reached. This process can be done for each object and for each pair ($h_{warp,max}$,$i$) to find a fit. 
In Section~3, we present the plot for the case $i=77^\circ$, and in Table~2 we show the value of $\delta$ obtained for each pair ($h_{warp,max}$,$i$). This table shows the degeneracy of the modeling and because the physical processes shaping the disk are not fully known, we cannot favor one solution over another. Our aim is to find the order of magnitude of $\delta$ such that we can conclude something about the degree of asymmetry of the inner disk.
In order to break the degeneracy, detailed MHD simulations including radiative transfer should be done and compared with resolved observations of the inner part of the disk to extract a value for $\delta$. and relate it to the actual disk height. Note that observationally $i$ can be fixed such that the modeling can be restricted to such a case. The parameter $\phi_{c}$ determines the azimuthal range where $h_{warp}=h_{min}$, such that it is related to the shape of the CoRoT lightcurve. A range of $\phi_{c}$ are given by McGinnis et al. (2015) in the modeling of the lightcurves of
the systems analyzed here. The value used here is within this range and fixed to the value $\phi_{c}=180$deg. A study changing this parameter is worth when the goal is to explain the details of the lightcurves. The pursuit of this requires a full understandings of all the physical processes involved based on a complete sample of hydrodynamical simulations that requires a gigantic amount of computational resources which is not what we are trying in this work.

\section{Lightcurve modeling}
\label{results}

\subsection{Modeling of Mon-660}
\label{Mon-660}

	This object was previously known as V354 Mon. Fonseca et al. (2014) interpreted its CoRoT lightcurve as due to occultations by an optically thick warp with a sinusoidal shape as the model Bouvier et al. (1999) used for the interpretation of the lightcurve of AA Tau. Observations in 2008 and 2011 are consistent with a stellar rotational period of $P_{rot}=5.25$days. The keplerian radius consistent with this period is $R_{k}=7.64R_{\star}$. Note that an explanation for the timescale of the variability requires that the material responsible for it should be located around this location. Remember that a bending wave located between $R_{k}=R_{cr}$ and $R_{ovr}$ moves with this periodicity, thus, it can be responsible for the stellar occultation and IR emission.

The stellar parameters are given in Table~1. The minimum sublimation radius is $R_{sub,min}=7.70R_{\star}$ using $\dot{M}=3\times 10^{-9}M_{\odot}yr^{-1}$. This value corresponds to the mean of $log(\dot{M})$ taken from Venuti et al. (2014). Note that $R_{sub,min}$ is slighly larger than $R_{k}$, because of the uncertainties we assume that the dusty wall is located at $R_{k}$.  

As mentioned in Section~2.3, Table~A1 present the set of pairs ($i$,$h_{max}$) explaining the observed $\Delta [CoRoT]\sim 0.8$. The value of $\delta$ that allows to explain the observed $\Delta [4.5]\sim 0.3$mag (McGinnis et al. 2015) for each pair and the mean observed flux at $4.5\mu$m ($<F_{obs}>$) are given in Table~2. The observed values for $\Delta [4.5]$ and $\Delta [CoRoT]$ are the maxima. 

In figure~3 we present the modeled and observed lightcurves for $i=77^\circ$ with $H_{min}=0.7R_{\star}$. We calculate a mean of the lightcurves, adding all the photometric cycles, such that all the observed points are used. We give the standard deviation $\sigma$ as a measure of the dispersion of the data, which is simply an effect of the variability of the lightcurves with respect to the mean curve. As a reference for the maximum amplitude observed, we plot the $+1\sigma$ lightcurves and fit their amplitudes.
The flux contributions from each component are presented in Figure~4. In the optical, the contribution from the wall and from the disk can be neglected, thus it is not showed.

   \begin{figure}
   \centering
   \includegraphics[angle=-90,width=\hsize]{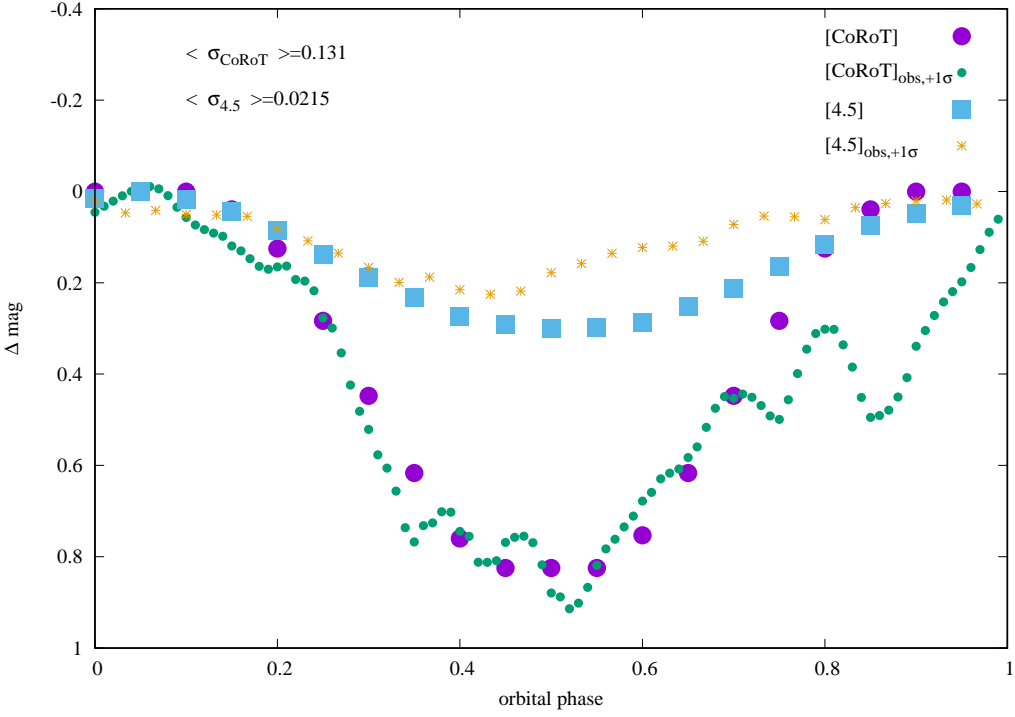}
      \caption{Modeled optical and $4.5\mu$m lightcurves for Mon-660 at $i=77^{\circ}$. The modeled CoRoT lightcurve is represented by filled circles, and the modeled Spitzer photometric magnitudes at $4.5\mu$m is plotted as solid squares. For comparison the observed $+1\sigma$ CoRoT (points) and the $+1\sigma$ Spitzer $4.5\mu$m (asterixs) lightcurves are presented. We use all the cycles presented in McGinnis et al. (2015) to calculate the observed curves. At the upper left corner, we show the value for the mean standard deviation for the CoRoT and Spitzer data. }
         \label{lc-Mon660}
   \end{figure}

   \begin{figure}
   \centering
   \includegraphics[angle=-90,width=\hsize]{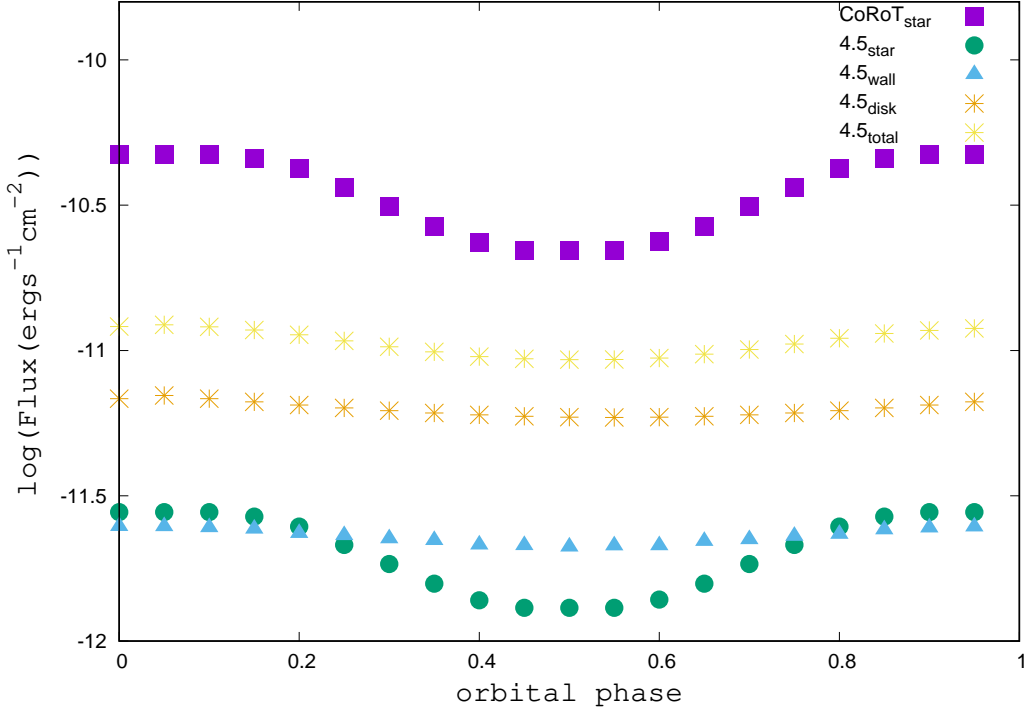}
      \caption{Flux contributions in the optical and the IR for Mon-660 at $i=77^{\circ}$. The modeled CoRoT and $4.5\mu$m fluxes coming from the star are represented as filled squares and filled circles, respectively. The $4.5\mu$m fluxes coming from the wall and the disk are plotted with solid triangles and asterixs, respectively. The light asterixs represent the total modeled flux.  }
         \label{Fluxes-Mon660}
   \end{figure}
 
\begin{table}
\caption{Stellar parameters}             
\label{table:parameters}      
\centering                          
\begin{tabular}{c c c c}        
\hline\hline                 
Object & $R_{\star}(R_{\odot})$ & $M_{\star}(M_{\odot})$ & $T_{\star}(K)$ \\     
\hline                        
 Mon-660  & 1.86 & 1.4 & 4574 \\      
 Mon-811 & 1.97 & 0.91 & 4196 \\
 Mon-1140 & 1.67 & 1.31 & 4578 \\
 Mon-1308 & 1.59 & 0.63 & 3909 \\
\hline                                   
\end{tabular}
\tablefoot{The first column shows the object name. The second column corresponds to the stellar radius. The third column is the stellar mass. The fourth column is the stellar effective temperature.}
\end{table}

\subsection{Modeling of Mon-811}
\label{Mon-811}

	From observations in 2011, $P_{rot}=7.88$days. The keplerian radius consistent with this period is $R_{k}=8.19R_{\star}$. 
The stellar parameters are given in Table~1. The minimum sublimation radius is $R_{sub,min}=6.19R_{\star}$ using $\dot{M}=3\times 10^{-9}M_{\odot}yr^{-1}$. This value corresponds to the mean of $log(\dot{M})$ taken from Venuti et al. (2014). 	

As mentioned in Section~2.3, Table~A1 present the set of pairs ($i$,$h_{max}$) explaining the observed $\Delta [CoRoT]=0.5$. The value of $\delta$ that allows to explain the observed $\Delta [4.5]\sim 0.2$mag (McGinnis et al. 2015) for each pair and $<F_{obs}>$ are given in Table~2. In figure~5 we present the modeled and observed lightcurves for $i=77^\circ$ with $H_{min}=0.8R_{\star}$. The flux contributions from each component are presented in Figure~6.

   \begin{figure}
   \centering
   \includegraphics[angle=-90,width=\hsize]{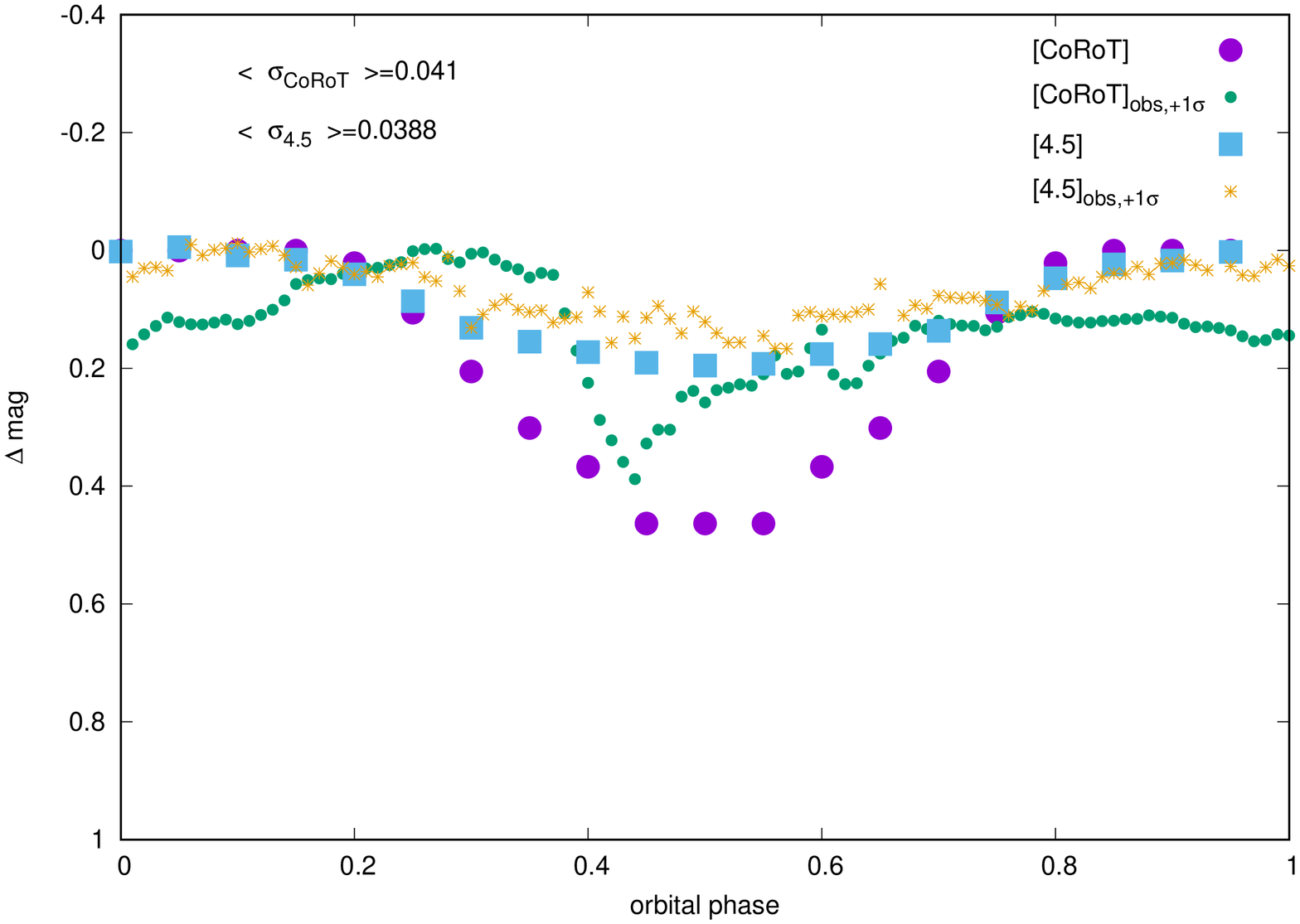}
      \caption{Modeled optical and IR lightcurves for Mon-811 at $i=77^{\circ}$. The symbols definitions are the same as in Figure~3.}
         \label{lc-Mon811}
   \end{figure}

   \begin{figure}
   \centering
   \includegraphics[angle=-90,width=\hsize]{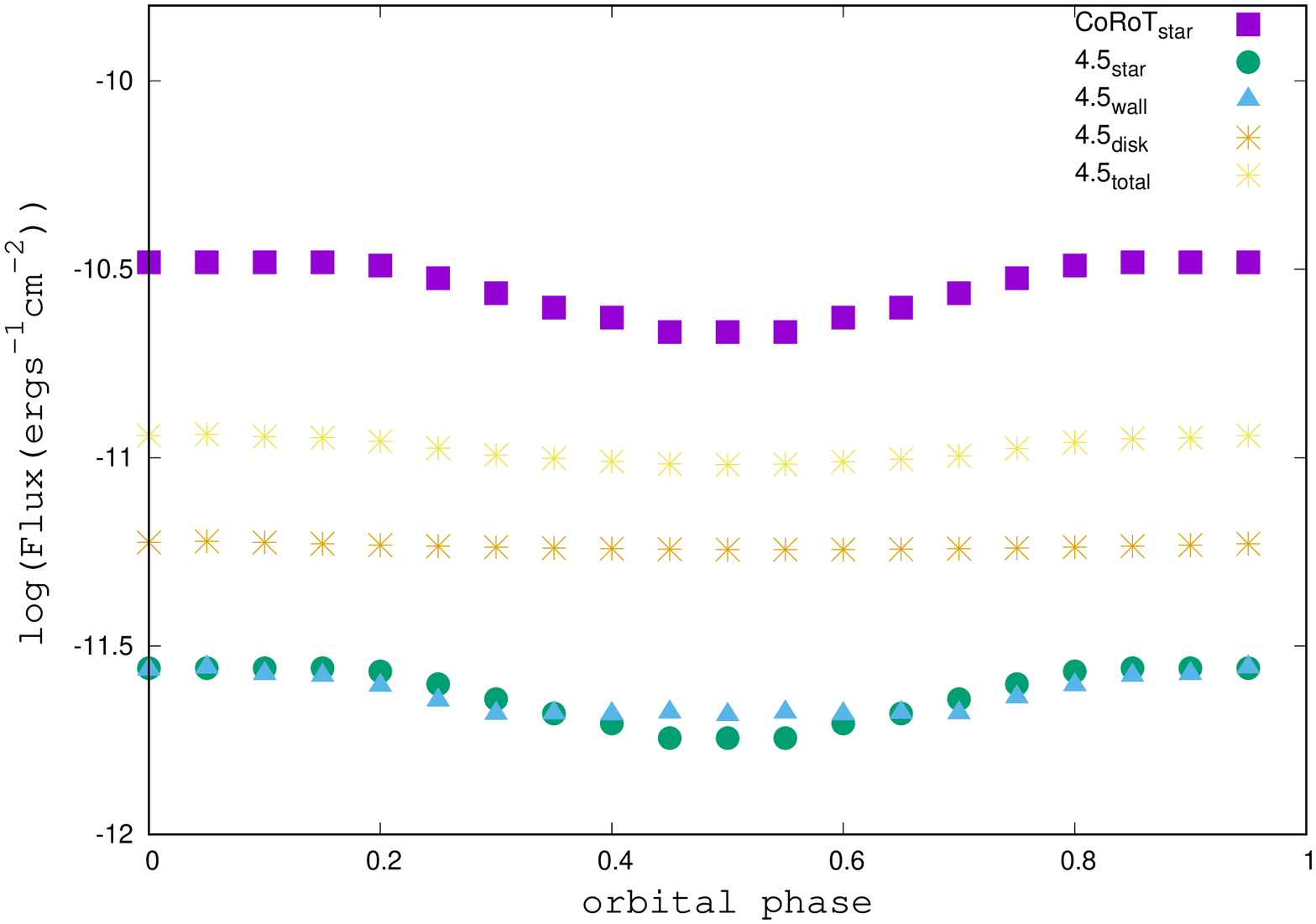}
      \caption{Flux contributions in the optical and the IR for Mon-811 at $i=77^{\circ}$. The symbols definitions are the same as in Figure~4.}
         \label{Fluxes-Mon811}
   \end{figure}

\begin{table}
\caption{Parameters in the models}          
\label{table:param-model}      
\centering                          
\begin{tabular}{c c c c c c c}        
\hline\hline                 
Object & $\dot{M}(M_{\odot}yr^{-1})$ & $R_{sub,min}(R_{\star})$  & $R_{k}(R_{\star})$ & $<F_{obs}>(erg\,cm^{-2}\,s^{-1})$ & $i(deg)$ & $\delta$ \\     
\hline                        
 Mon-660  & $3\times 10^{-9}$ & $7.7$ & $7.64$ & $9.51\times 10^{-12}$ &
 71 & 0.15  \\     
 Mon-660  & $3\times 10^{-9}$ & $7.7$ & $7.64$ & $9.51\times 10^{-12}$ &
 72 & 0.15  \\      
 Mon-660  & $3\times 10^{-9}$ & $7.7$ & $7.64$ & $9.51\times 10^{-12}$ &
 73 & 0.17  \\      
 Mon-660  & $3\times 10^{-9}$ & $7.7$ & $7.64$ & $9.51\times 10^{-12}$ &
 74 & 0.15  \\      
 Mon-660  & $3\times 10^{-9}$ & $7.7$ & $7.64$ & $9.51\times 10^{-12}$ &
 75 & 0.14  \\      
 Mon-660  & $3\times 10^{-9}$ & $7.7$ & $7.64$ & $9.51\times 10^{-12}$ &
 76 & 0.15  \\      
 Mon-660  & $3\times 10^{-9}$ & $7.7$ & $7.64$ & $9.51\times 10^{-12}$ &
 77 & 0.16  \\      
\hline
 Mon-811  & $3\times 10^{-9}$ & $6.19$ & $8.19$ & $8.65\times 10^{-12}$ &
 67 & 0.03  \\     
 Mon-811  & $3\times 10^{-9}$ & $6.19$ & $8.19$ & $8.65\times 10^{-12}$ &
 68 & 0.02  \\     
 Mon-811  & $3\times 10^{-9}$ & $6.19$ & $8.19$ & $8.65\times 10^{-12}$ &
 69 & 0.0  \\     
 Mon-811  & $3\times 10^{-9}$ & $6.19$ & $8.19$ & $8.65\times 10^{-12}$ &
 70 & 0.0  \\     
 Mon-811  & $3\times 10^{-9}$ & $6.19$ & $8.19$ & $8.65\times 10^{-12}$ &
 71 & 0.02  \\     
 Mon-811  & $3\times 10^{-9}$ & $6.19$ & $8.19$ & $8.65\times 10^{-12}$ &
 72 & 0.03  \\     
 Mon-811  & $3\times 10^{-9}$ & $6.19$ & $8.19$ & $8.65\times 10^{-12}$ &
 73 & 0.03  \\     
 Mon-811  & $3\times 10^{-9}$ & $6.19$ & $8.19$ & $8.65\times 10^{-12}$ &
 74 & 0.03  \\     
 Mon-811  & $3\times 10^{-9}$ & $6.19$ & $8.19$ & $8.65\times 10^{-12}$ &
 75 & 0.01  \\     
 Mon-811  & $3\times 10^{-9}$ & $6.19$ & $8.19$ & $8.65\times 10^{-12}$ &
 76 & 0.02  \\     
 Mon-811  & $3\times 10^{-9}$ & $6.19$ & $8.19$ & $8.65\times 10^{-12}$ &
 77 & 0.05  \\     
\hline
 Mon-1140  & $7.76\times 10^{-9}$ & $7.93$ & $6.75$ & $3.76\times 10^{-12}$ &
 73-77 & 0.01  \\     
\hline
 Mon-1308  & $8.51\times 10^{-9}$ & $5.45$ & $7.81$ & $2.77\times 10^{-12}$ &
 69 & 0.2  \\     
 Mon-1308  & $8.51\times 10^{-9}$ & $5.45$ & $7.81$ & $2.77\times 10^{-12}$ &
 70 & 0.25  \\     
 Mon-1308  & $8.51\times 10^{-9}$ & $5.45$ & $7.81$ & $2.77\times 10^{-12}$ &
 71 & 0.4  \\     
 Mon-1308  & $8.51\times 10^{-9}$ & $5.45$ & $7.81$ & $2.77\times 10^{-12}$ &
 72 & 0.45  \\     
 Mon-1308  & $8.51\times 10^{-9}$ & $5.45$ & $7.81$ & $2.77\times 10^{-12}$ &
 73 & 0.35  \\     
 Mon-1308  & $8.51\times 10^{-9}$ & $5.45$ & $7.81$ & $2.77\times 10^{-12}$ &
 74 & 0.38  \\     
 Mon-1308  & $8.51\times 10^{-9}$ & $5.45$ & $7.81$ & $2.77\times 10^{-12}$ &
 75 & 0.4  \\     
 Mon-1308  & $8.51\times 10^{-9}$ & $5.45$ & $7.81$ & $2.77\times 10^{-12}$ &
 76 & 0.4  \\     
 Mon-1308  & $8.51\times 10^{-9}$ & $5.45$ & $7.81$ & $2.77\times 10^{-12}$ &
 77 & 0.56  \\     
\hline                                   
\end{tabular}
\tablefoot{The first column shows the object name. The second column corresponds to the mass accretion rate. The third column is the minimum sublimation radius. The forth column is the keplerian radius. The fifth column is the observed flux at $4.5\mu$m. The sixth column corresponds to the inclination. The seventh column is the 
$\delta$ parameter (see text for details).}
\end{table}

\subsection{Modeling of Mon-1140}
\label{Mon-1140}

	From observations in 2008 and 2011, $P_{rot}=3.87$ and $3.9$days, respectively. The keplerian radius consistent with the latter period is $R_{k}=6.75R_{\star}$. 
The stellar parameters are given in Table~1. The minimum sublimation radius is $R_{sub,min}=7.93R_{\star}$ using $\dot{M}=7.76\times 10^{-9}M_{\odot}yr^{-1}$. The value for $\dot{M}$ is taken from Venuti et al. (2014). They present two different estimates of $\dot{M}$ and we decided to take the value with the lowest $R_{sub,min}$. However, even in this case $R_{sub,min}> R_{k}$ and we can conclude that there is not dust at $R_{k}=R_{mag}$, such that there is not a wall formed by magnetospheric streams. For this object, the bending wave and/or material above it is responsible for the stellar occultation. Noteworthy, the value for $<F_{disk}>$
calculated using equation~4 is a few times smaller than required to explain the observed $\Delta [4.5]$. Thus, for this object the IR lightcurves are interpreted only using the emission coming from the disk, such that instead of equation~4, we use $<F_{obs}>=<F_{disk}>$.  	
The value of $\delta$ that allows to explain the observed $\Delta [4.5]\sim 0.1$mag (McGinnis et al. 2015) and $<F_{obs}>$ are given in Table~2. In Figure~7 we present the modeled and observed $[4.5]$ lightcurves. For the modeling of this object, a change in $i$ means that the physical configuration required to get $F_{disk}$ changes. The analysis of this changing disk configuration is not pursued in this work. The optical lightcurves presented correspond to a vertical wall located at $R_{sub,min}$ with the height given by $h_{warp}R_{sub,min}/R_{k}$. The geometrical characterization of the bending wave is not the aim of this work. The flux contributions from each component are presented in Figure~8.

   \begin{figure}
   \centering
   \includegraphics[angle=-90,width=\hsize]{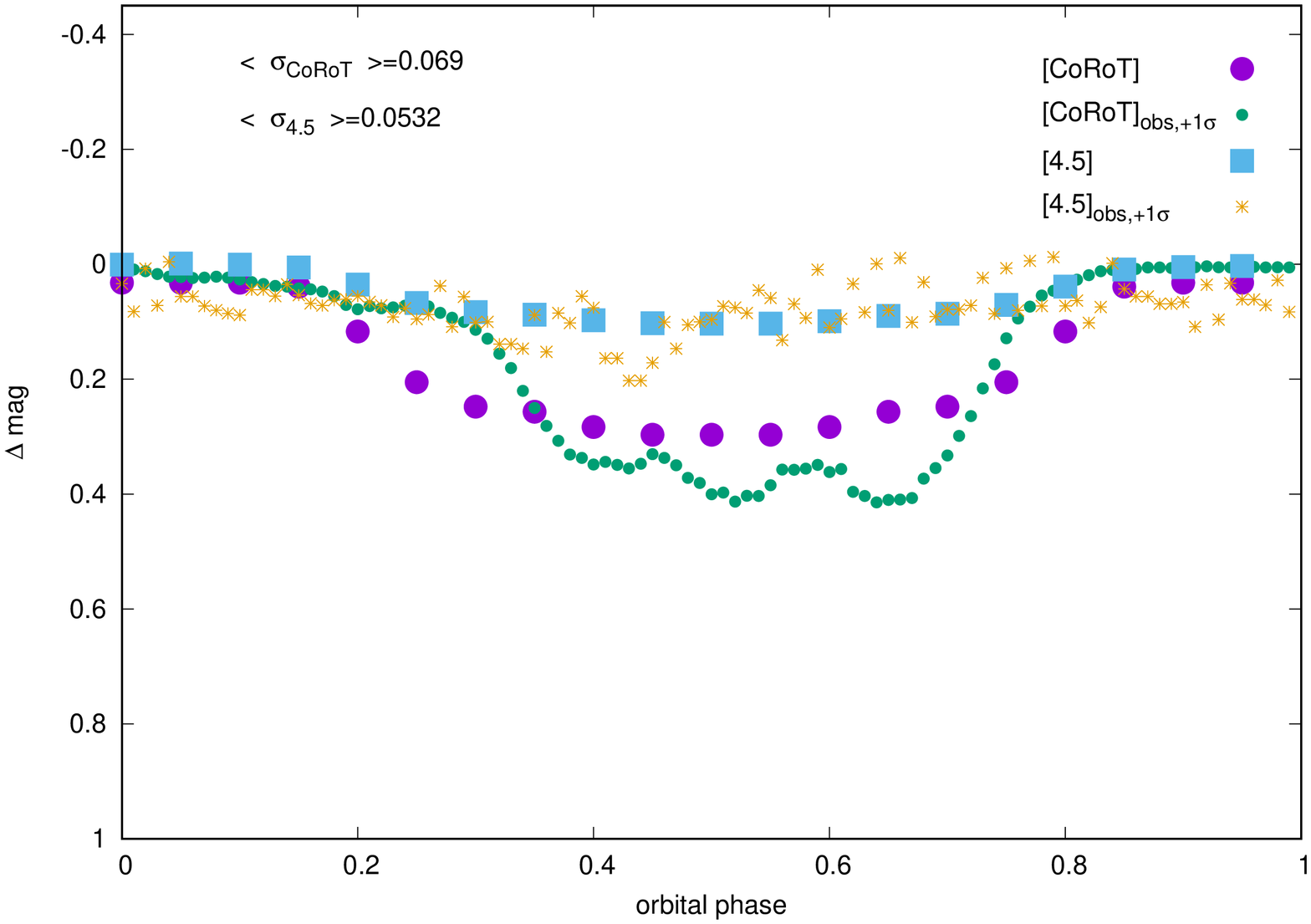}
      \caption{Modeled optical and IR lightcurves for Mon-1140. The symbols definitions are the same as in Figure~3.}
         \label{lc-Mon1140}
   \end{figure}
 
   \begin{figure}
   \centering
   \includegraphics[angle=-90,width=\hsize]{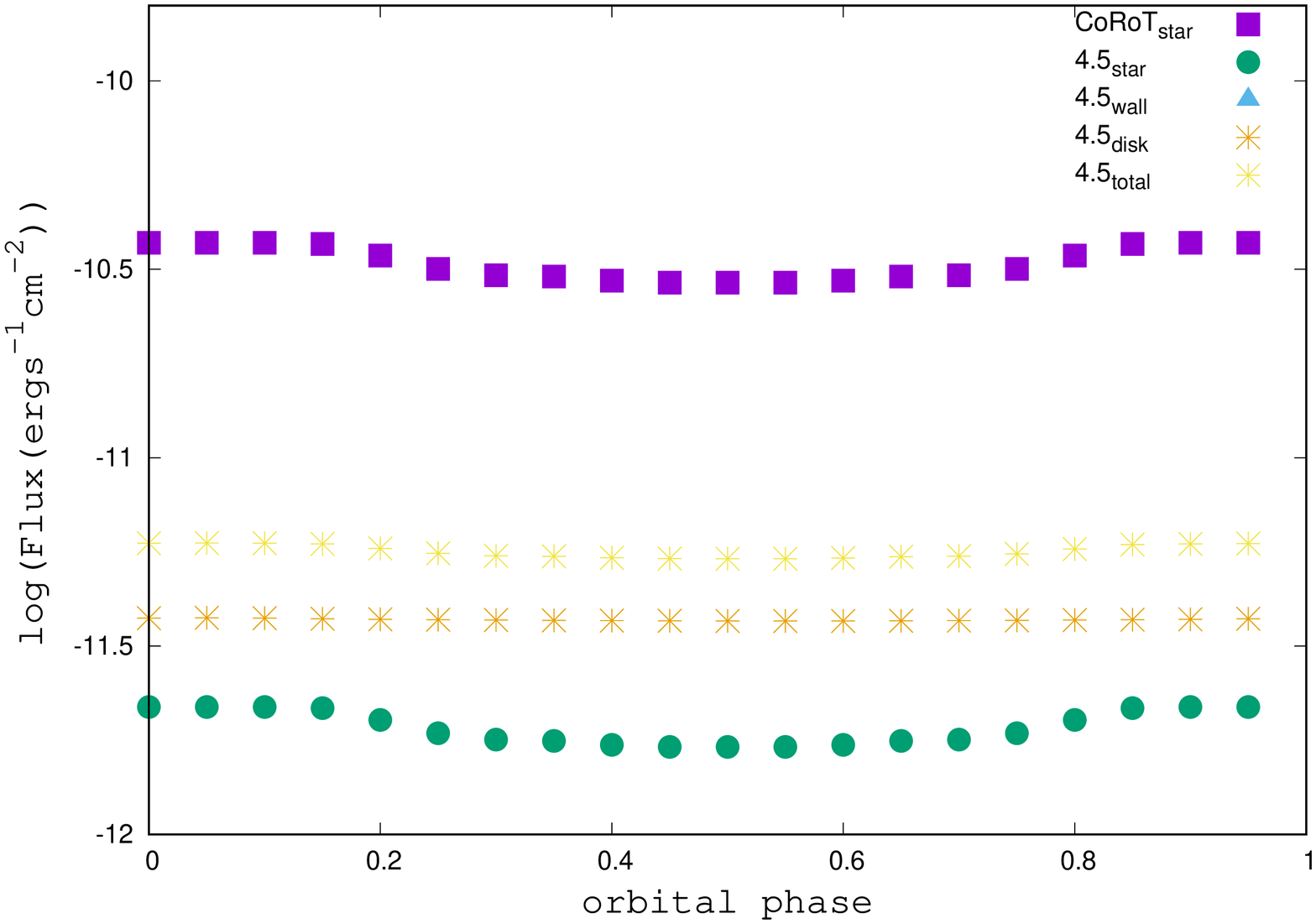}
      \caption{Flux contributions in the optical and the IR for Mon-1140. The symbols definitions are the same as in Figure~4.}
         \label{Fluxes-Mon1140}
   \end{figure}

\subsection{Modeling of Mon-1308}
\label{Mon-1308}

	From observations in 2008 and 2011, $P_{rot}=6.45$ and $6.68$days. The keplerian radius consistent with this period is $R_{k}=7.81R_{\star}$. 
The stellar parameters are given in Table~1. The minimum sublimation radius is $R_{sub,min}=5.45R_{\star}$ using $\dot{M}=8.51\times 10^{-9}M_{\odot}yr^{-1}$. The value for $\dot{M}$ is taken from Venuti et al. (2014).	

As mentioned in Section~2.3, Table~A1 present the set of pairs ($i$,$h_{max}$) explaining the observed $\Delta [CoRoT]=0.4$. The value of $\delta$ that allows to explain the observed $\Delta [4.5]\sim 0.4$mag (McGinnis et al. 2015) for each pair and $<F_{obs}>$ are given in Table~2. In Figure~9 we present the modeled and observed lightcurves for $i=77^\circ$ with $H_{min}=0.8R_{\star}$. The flux contributions from each component are presented in Figure~10.

   \begin{figure}
   \centering
   \includegraphics[angle=-90,width=\hsize]{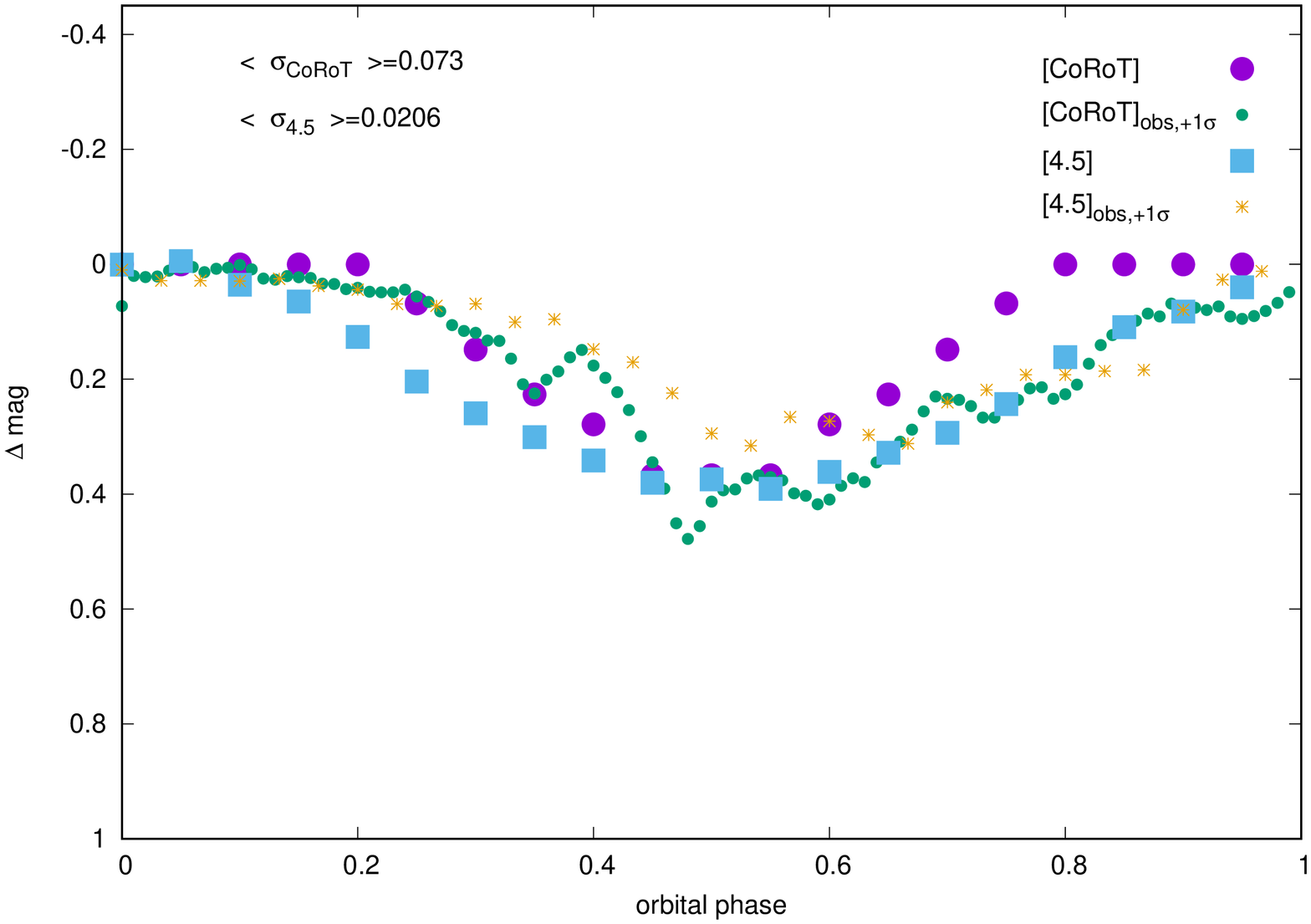}
      \caption{Modeled optical and IR lightcurves for Mon-1308 at $i=77^{\circ}$. The symbols definitions are the same as in Figure~3.}
         \label{lc-Mon1308}
   \end{figure}

   \begin{figure}
   \centering
   \includegraphics[angle=-90,width=\hsize]{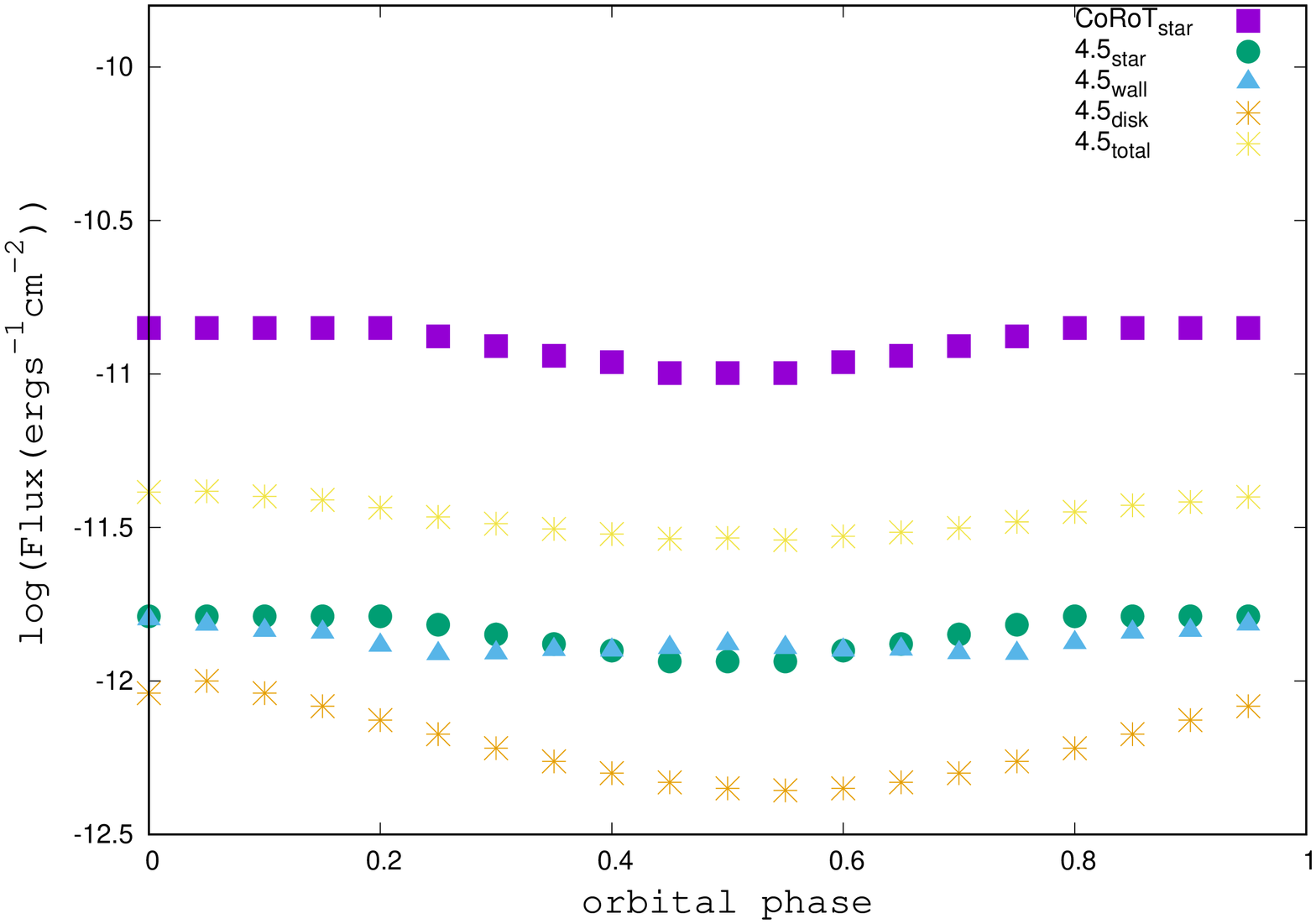}
      \caption{Flux contributions in the optical and the IR for Mon-1308 at $i=77^{\circ}$. The symbols definitions are the same as in Figure~4.}
         \label{Fluxes-Mon1308}
   \end{figure}

\section{Discussion}
\label{discussion}

The Spitzer lightcurves for the objects Mon-660, Mon-811, and Mon-1308 can be explained using the emission coming from the star, the vertical wall and the  asymmetrical emission associated to a disk behind the wall. The wall is located at the keplerian radius consistent with the period of the lightcurves. The asymmetry of this structure is the result of the interaction between a stellar magnetic field with a tilted dipolar axis and the disk. For Mon-1140, at the location of the wall, the stellar heating sublimates the dust such that the IR lightcurves are explained only with the stellar and the disk emission.  

The CoRoT lightcurves for Mon-660, Mon-811, and Mon-1308 are interpreted with occultation for a vertical wall, however, for Mon-1140 there is not a wall such that the occulting structure should lie in the asymmetrical disk. The MHD simulations in Romanova et al. (2013) suggest that vertical perturbations in the disk can be responsible for the stellar shadowing. The real fact is that dust should be moved upwards in order to periodically block a section of the stellar surface.  
 
Romanova et al. (2013) describe the resulting configuration for the interaction between the stellar magnetosphere and the disk as a bending wave that is located between the corotation radius ($R_{cr}$) and the location for the outer vertical resonance ($R_{ovr}=4^{1/3}R_{cr}$). Note that for rapidly rotating stars, $R_{cr}=R_{mag}$ is close to the rotational equilibrium state, which is a state commonly reached by accreting magnetized stars (Long et al. 2005). In our case, $R_{k}=R_{mag}$ such that $R_{k}=R_{cr}$. The wave located in this radial range moves at the stellar period such that any section of it can be responsible of the stellar occultations required to explain the CoRoT lightcurves. Because the emission of such structure is beyond the scope of this paper, we simply quantify its emission as the missing contribution required to explain the Spitzer lightcurves (see Section~2.2). 

The value for $\delta$ required to explain the Spitzer lightcurves is the largest for the systems with the largest $\Delta[4.5]$: $\Delta[4.5]=0.3$ for Mon-660 and $\Delta[4.5]=0.4$ for Mon-1308. This means that a larger asymmetry in the flux (and accordingly in the disk configuration) is required to interpret this kind of systems. For the systems with the lowest $\Delta[4.5]$
($\Delta[4.5]=0.2$ for Mon-811 and $\Delta[4.5]=0.1$ for Mon-1140), the value for $\delta$ required is the lowest, with values down to $\delta=0.01$. For this case, the disk does not show a large asymmetry, in other words, the vertical size of the wave is not large. The vertical size of the actual structures in the disk, $h_{disk}$, is directly related to the emitting area so that a larger $h_{disk}$ corresponds to a larger observed flux. Radiative MHD simulations are required to connect $\delta$ with $h_{disk}$. However an estimate of this relation can be done assuming two facts: 1) ${\delta F_{disk}\over F_{wall}}$ is the emission associated to the structure above the disk in units of $F_{wall}$ and 2) the wall emission is proportional to $h_{warp,max}$ such that the disk emission is proportional to the disk surface height $h_{disk}$. Using these assumptions, $h_{disk}={\delta F_{disk}\over F_{wall}}\times h_{warp,max}$. This can be applied to the 3 systems with a dusty wall,
  resulting in
$h_{disk}=(0.46,0.05,0.12)R_{\star}$ for Mon-660, 811 and 1308, respectively. These values can be compared to the model FW$\mu 1.5$ in Romanova et al. (2013), where the largest amplitude for the warp is $0.57R_{\star}$. For Mon-660, these values are comparable but for Mon-811 and Mon-1308, $h_{disk}$ is lower. This means that for the last two cases, the interaction of the dipolar stellar magnetic field with the disk is weaker. Radiative MHD simulations are required to test these estimates using more detailed physical input.  

For Mon-1140 it is satisfied $R_{sub,min}> R_{k}$, such that there is not dust at $R_{k}$. This means that all the lightcurves interpretation is based on the bending wave. A value of $\delta=0.01$ for this system means that $F_{disk}$ should change only $1\%$ in order to explain the flux variability.

\section{Conclusions}
\label{conclusions}

Our main conclusion is that an optically thick wall at the keplerian radius associated to the periodicity of the observed lightcurve and an asymmetric disk are able to consistently explain the CoRoT and the Spitzer lightcurves of the NGC 2264 dippers Mon-660, Mon-811, Mon-1140 and Mon-1308. The stellar occultation by the wall or the asymmetrical structure in the disk is responsible for the modulation of the optical lightcurve and the emission from the partially occulted star, and the optically thick warp can explain the IR lightcurve.

A more detailed analysis of the effects of the distribution of material in the warp on the modeling will require to run hydrodynamical simulations to get the distribution of gas and dust in the 3-D structure. However, the simple structure considered here is enough to justify the basic picture to explain the lightcurves. 

\begin{acknowledgements}
      E.N. appreciate the support from the Institut de Plan\'etologie et d'Astrophysique de Grenoble (Universit\'e Grenoble Alpes) during a sabbatical stay where most of the work has been done.	
      This project has received funding from the European Research Council (ERC) under the European Union's Horizon 2020 research and innovation programme (grant agreement No 742095; {\it SPIDI}: Star-Planets-Inner
Disk-Interactions)	
\end{acknowledgements}

\begin{appendix} 
\section{Appendix: Degeneracy between i and hwarp,max}
\label{table}

There is a geometrical degeneracy between the parameters $i$ and $h_{warp,max}$ because a larger $i$ means that a lower $h_{warp,max}$ is required to get the amount of the occultation necessary to explain the $[CoRoT]$ lightcurve. In Table~A1 we put the pairs of values producing the same amplitude for the observed $\Delta [CoRoT]$. For each object, the $i$ range is defined with the estimated value and its error obtained in the previous modeling of McGinnis et al. (2015). They obtained projected rotational velocities ($vsini$) comparing FLAMES or/and Hectochelle spectra to synthetic spectra and then they found $i$ using the relation $vsini={2\pi R_{star}\over P_{rot}}sini$. The maximum value of $i$ is fixed assuming a typical flared disk, such that the star is not completely occulted by the disk.


    \begin{table}	
      \caption[]{$i$ and $h_{warp,max}$ consistent with observed
$\Delta [CoRoT]$ for each object.}
         \label{table:i-vs-hmax}
    \centering	
    \begin{tabular}{c c c c c c c c c c c c}
\hline\hline                 
            \hline
            Mon-660 & \multicolumn{11}{c}{}  \\
            \hline
  $i(deg)$ &... &... &... &... & 71 & 72 & 73 & 74 & 75 & 76 & 77 \\
  $h_{warp,max} (R_{\star})$ &... &... & ... &...& 1.9 & 1.75 & 1.6 & 1.5 & 1.4 & 1.2 & 1.1 \\
  \hline  
  Mon-811 & & & & & & & & & & & \\
  \hline	
  $i(deg)$ & 67 & 68 & 69 & 70 & 71 & 72 & 73 & 74 & 75 & 76 & 77 \\
  $h_{warp,max} (R_{\star})$ & 2.0 & 1.9 & 1.8 & 1.6 & 1.5 & 1.4 & 1.2 & 1.1 & 1.0 & 0.8 & 0.7\\
  \hline  
  Mon-1140 & & & & & & & & & & & \\
  \hline	
  $i(deg)$ & ...& ...& ...& ...& ... & ...& 73 & 74 & 75 & 76 & 77 \\
  $h_{warp,max} (R_{\star})$ & ...& ...& ...& ...& ... & ...& 1.1 & 0.9 & 0.8 & 0.7 & 0.6 \\
  \hline  
  Mon-1308 & & & & & & & & & & & \\
  \hline	
  $i(deg)$ & ...& ...& 69 & 70 & 71 & 72 & 73 & 74 & 75 & 76 & 77 \\
  $h_{warp,max} (R_{\star})$ & ...& ...& 1.7 & 1.5 & 1.4 & 1.3 & 1.1 & 1.0 & 0.9 & 0.7 & 0.6 \\
  \hline  
   \end{tabular}
   \end{table}		


\end{appendix}

\end{document}